\documentclass[twocolumn,showpacs,preprintnumbers,amsmath,amssymb,prl]{revtex4}

\usepackage{graphicx}
\usepackage{bm}

\begin{document}

\title{Water-like Anomalies and Breakdown of the Rosenfeld Excess Entropy
Scaling Relations for the Core-Softened Systems:
Dependence on the Trajectory in Density-Temperature Plane}

\author{Yu. D. Fomin}
\affiliation{Institute for High Pressure Physics, Russian Academy
of Sciences, Troitsk 142190, Moscow Region, Russia}

\author{V. N. Ryzhov}
\affiliation{Institute for High Pressure Physics, Russian Academy
of Sciences, Troitsk 142190, Moscow Region, Russia}

\date{\today}

\begin{abstract}
We show that the existence of the water-like anomalies in kinetic
coefficients in the core-softened systems depends on the
trajectory in $\rho-T$ plane along which the kinetic coefficients
are calculated. In particular, it is shown that the diffusion
anomaly does exist along the isotherms, but disappears along the
isochores. We analyze the applicability of the Rosenfeld entropy
scaling relations to the systems with the core-softened potentials
demonstrating the water-like anomalies. It is shown that the
validity of the of Rosenfeld scaling relation for the diffusion
coefficient also depends on the trajectory in the $\rho-T$ plane
along which the kinetic coefficients and the excess entropy are
calculated. In particular, it is valid along isochors, but it
breaks down along isotherms.
\end{abstract}

 \maketitle

It is well known that some liquids (for example, water, silica,
silicon, carbon, and phosphorus) show an anomalous behavior \cite{
book,book1,deben2001,netz,errington1,errington2}:their phase
diagrams have regions where a thermal expansion coefficient is
negative (density anomaly), self-diffusivity increases upon
compression (diffusion anomaly), and the structural order of the
system decreases with increasing pressure (structural anomaly)
\cite{deben2001,netz}. A number of studies demonstrate water-like
anomalies in fluids that interact through spherically symmetric
potentials (see, for example, \cite{buld2009,weros,wepre} and
references therein).

It was shown \cite{errington1,errington2} that thermodynamic and
kinetic anomalies may be linked through the excess entropy. In
particular, in Refs. [\onlinecite{errington1,errington2}] the
authors propose that the entropy scaling relations developed by
Rosenfeld can be used to describe the diffusivity anomaly.

In 1977 Rosenfeld proposed the relations connecting transport
properties of a liquid with the excess entropy \cite{ros1}. In
order to write down these relations one should use the reduced
forms of the transport coefficients:
\begin{equation}
 D^*=D \frac{\rho ^{1/3}}{(k_BT/m)^{1/2}}, \label{1}
\end{equation}
where $D$ is the diffusion coefficient. According to the Rosenfeld
suggestion, the reduced diffusion can be expressed in the form
\begin{equation}
  D^*=a \cdot e^{b S_{ex}}, \label{3}
\end{equation}
where $S_{ex}=(S-S_{id})/{(Nk_B)}$ is excess entropy of the liquid
and $a$ and $b$ are the constants which depend on the studying
property \cite{ros2}. The coefficients $a$ and $b$ show an
extremely weak dependence on the material and can be considered as
universal.

In his original works Rosenfeld considered hard spheres, soft
spheres, Lennard-Jones system and one-component plasma
\cite{ros1,ros2}. After that the excess entropy scaling was
applied to many different systems including core-softened liquids
\cite{errington1,errington2,india1,indiabarb,weros}, liquid metals
\cite{liqmet1,liqmet2}, binary mixtures \cite{binary1,binary2},
ionic liquids \cite{india2,ionicmelts}, network-forming liquids
\cite{india1,india2}, water \cite{buldwater}, chain fluids
\cite{chainfluids} and bounded potentials
\cite{weros,klekelberg,klekelberg1}.

Nevertheless, controversies still remain. For example, up to the
moment it is not clear whether the Rosenfeld scaling relations are
applicable to the core softened systems. Some publications state
that the scaling relations are valid for such systems
\cite{errington2,indiabarb}, while in our recent work it was shown
that the scaling relations may break down for the core softened
systems \cite{weros}. This article presents a discussion of this
contradiction. Basing on the molecular dynamics simulations of the
two core-softened systems we show that the existence of the
water-like anomalies and the validity of the Rosenfeld scaling
relations depend on the trajectory in the ($\rho-T$) plane, along
which the kinetic coefficients are calculated. For the first time
it is explicitly shown that the water-like anomalies in the
systems with core-softened potentials do exists when the kinetic
coefficients are calculated along the isotherms and do not exist
along the isochores. Consequently, the exponential functional form
of the relation between the excess entropy and the reduced
diffusion coefficient holds along isochors while along isotherms
one observes its breakdown.

Two systems are studied in the present work. The first one is a
repulsive shoulder system (RSS) introduced in our previous works
\cite{wejcp,wepre}. This system has a potential
\begin{equation}
  U(r)=
  \left(\frac{\sigma}{r}\right)^{14}+\frac{1}{2}\varepsilon
  \cdot[1-\tanh(k_0\{r-\sigma_1\})],
\end{equation}
where $\sigma$ is the "hard"-core diameter, $\sigma_1=1.35$ is the
soft-core diameter, and $k_0=10.0$. In Ref. \cite{wepre} it was
shown that this system demonstrates anomalous thermodynamic
behavior. In our previous publication \cite{weros} the Rosenfeld
relation for this system was studied. It was shown that the
scaling relation for the diffusion coefficient breaks down for
this system in the anomalous diffusion region along the isotherms.

The second system studied in this work is the core-softened system
introduced by de Oliveira et al \cite{barbosapot}. This system is
described by the spherically symmetric potential represented by a
sum of a Lennard-Jones contribution and a Gaussian-core
interaction (LJG):
\begin{equation}
  U(r)=4\varepsilon
  \left[\left(\frac{\sigma}{r}\right)^{12}-\left(\frac{\sigma}{r}\right)^{6}\right]
  +a\varepsilon
  \cdot \exp\left[-\frac{1}{c^2}\left(\frac{r-r_0}{\sigma}\right)^2\right],
\end{equation}
with $a=5.0$, $r_0/ \sigma=0.7$ and $c=1.0$. This model can
qualitatively reproduce water's density, diffusivity, and
structural anomalies. The diffusivity of this system was studied,
for example, in the papers \cite{barbosapot,indiabarb}.

In this paper we use the dimensionless quantities: $\tilde{{\bf
r}}={\bf r}/ \sigma$, $\tilde{P}=P \sigma
^{3}/\varepsilon ,$ $\tilde{V}=V/N \sigma^{3}=1/\tilde{\rho},$ $\tilde{T}%
=k_{B}T/\varepsilon $. Since we use only these reduced units we
omit the tilde marks.

The simulation setup of the RSS was described in detail in Ref.
\cite{weros}. The following isotherms are simulated: $T=0.2; 0.25;
0.3; 0.35; 0.4; 0.5; 0.6; 0.7$ and $0.8$.

For the investigation of the LJG potential we simulate a system of
$1000$ particles in a cubic box for the densities ranging from
$\rho=0.01$ till $\rho=0.35$ with the step $\delta \rho =0.01$.
The time step used is $dt=0.001$. The equilibration period
consists of $1 \cdot 10^6$ time steps and the production period -
$2.5 \cdot 10^6$ time steps. During the equilibration the
temperature is kept constant by velocity rescaling while during
the production cycle $NVE$-MD is used. The equations of motion are
integrated by velocity-Verlet algorithm. The following isotherms
are simulated: $T=0.2; 0.3; 0.4; 0.5; 0.6; 1.0$ and $1.5$.

The excess entropy in both cases was computed via thermodynamic
integration method. For doing this we calculate excess free energy
by integrating the equation of state along an isotherm:
$\frac{F_{ex}}{Nk_BT}=\frac{F-F_{id}}{Nk_BT}=\frac{1}{k_BT}
\int_0^{\rho} \frac{P(\rho ')-\rho ' k_BT}{\rho '^2} d\rho'$. The
excess entropy is computed via $S_{ex}=\frac{U-F_{ex}}{N k_BT}$.


\begin{figure}
\includegraphics[width=5cm, height=4cm]{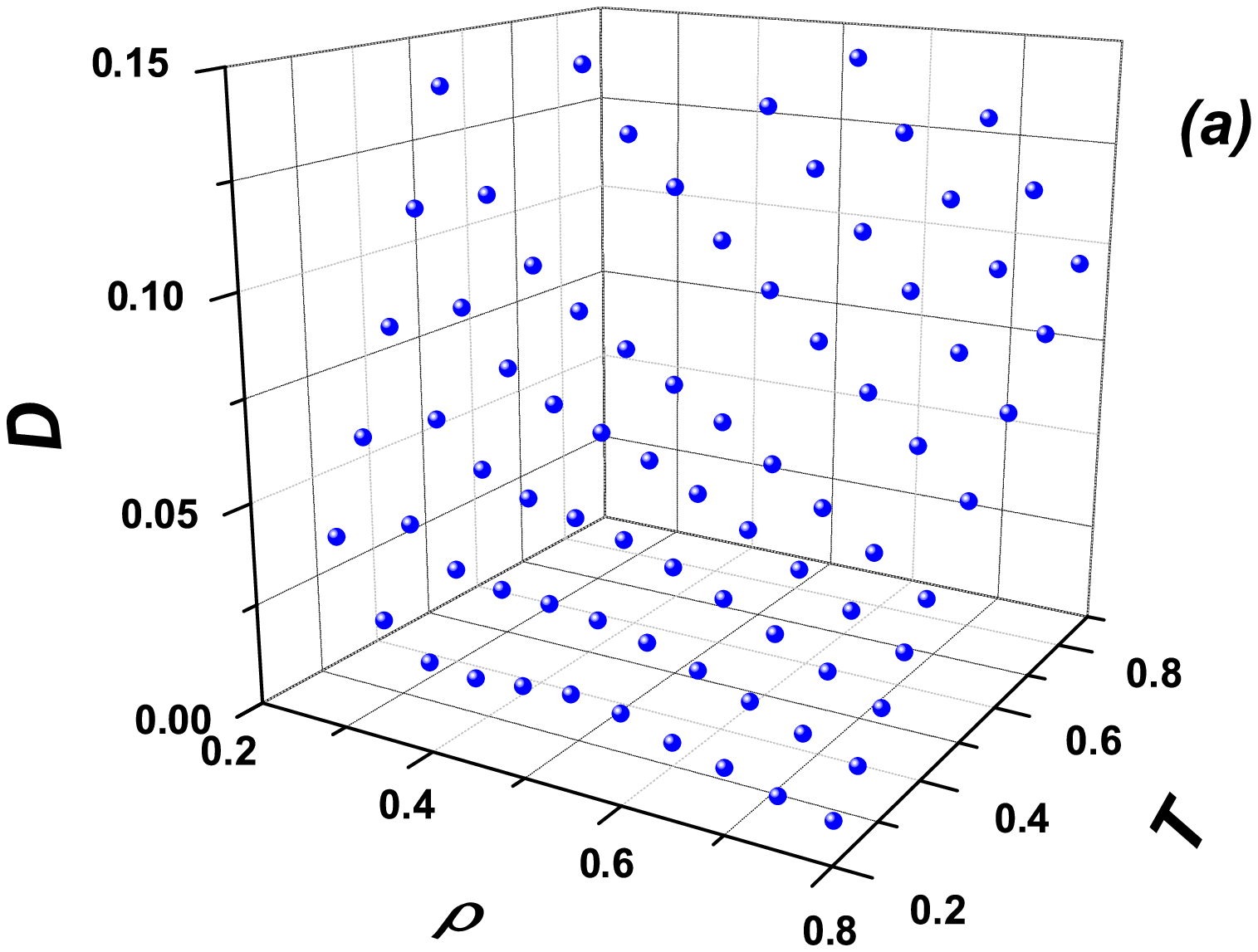}%
\bigskip
\bigskip

\includegraphics[width=5cm, height=4cm]{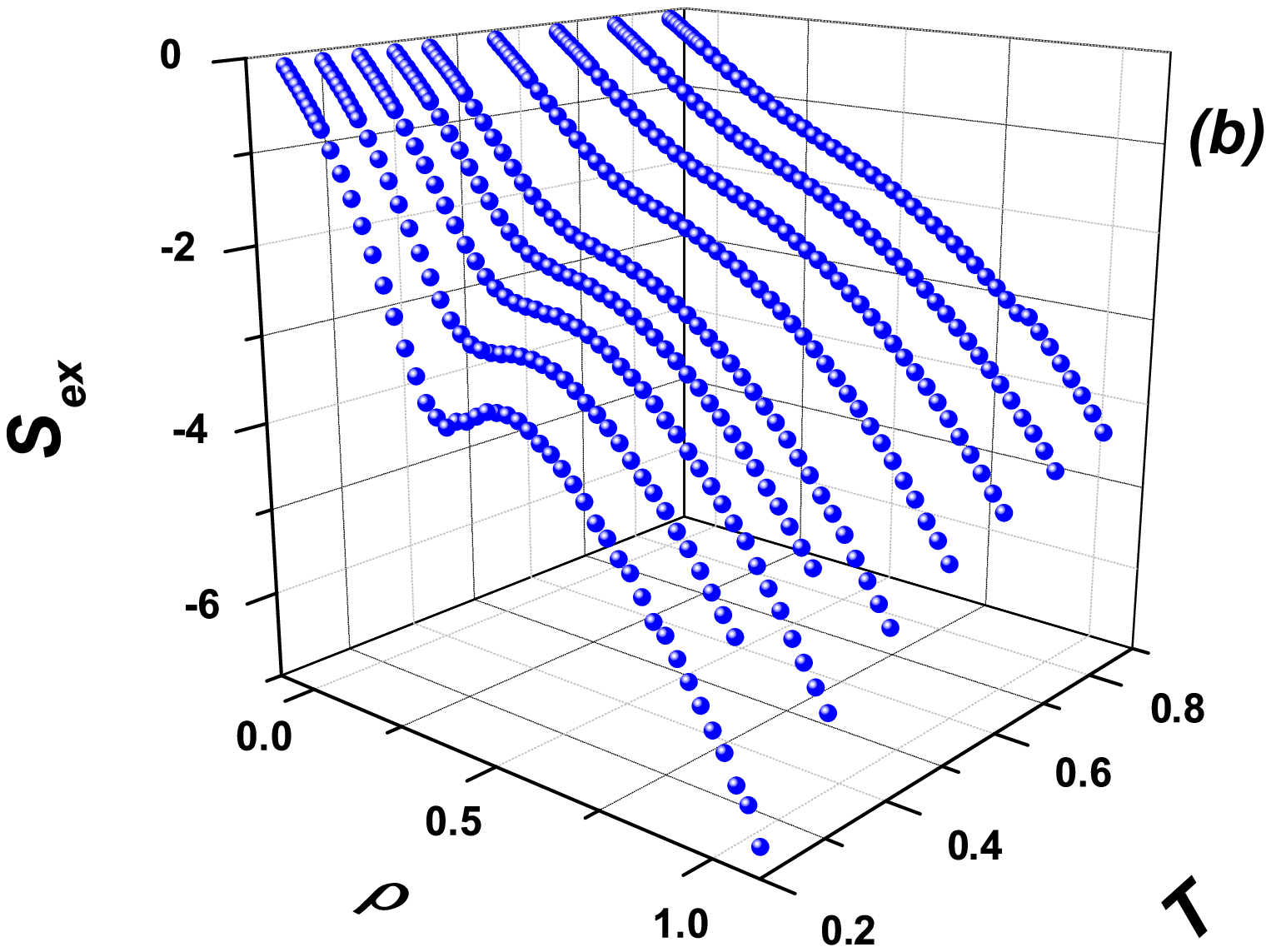}%
\bigskip

\caption{\label{fig:fig1} (Color online) (a) The diffusion
coefficient of the RSS system as a function of $\rho$ and $T$. (b)
The excess entropy of the RSS system as a function of $\rho$ and
$T$. }
\end{figure}

Fig.~\ref{fig:fig1}(a) presents the diffusion coefficient of the
RSS system as a function of $\rho$ and $T$. Fig.~\ref{fig:fig1}(b)
demonstrates the excess entropy as a function of $\rho$ and $T$.
From these figures it follows that the diffusion coefficient and
the excess entropy along an isotherm have similar qualitative
behavior. From Figs.~\ref{fig:fig1}(a) one can see that there is a
diffusion anomaly along the isotherms, but there is no anomaly
along the isochores. Moreover, it is possible to find the
trajectories in the ($\rho-T$) plane, along which the diffusion
anomaly does not exist. For example, it may be shown that there is
no diffusion anomaly along the isobars. This issue will be
discussed in detail in the subsequent publication.

\begin{figure}
\includegraphics[width=6cm, height=6cm]{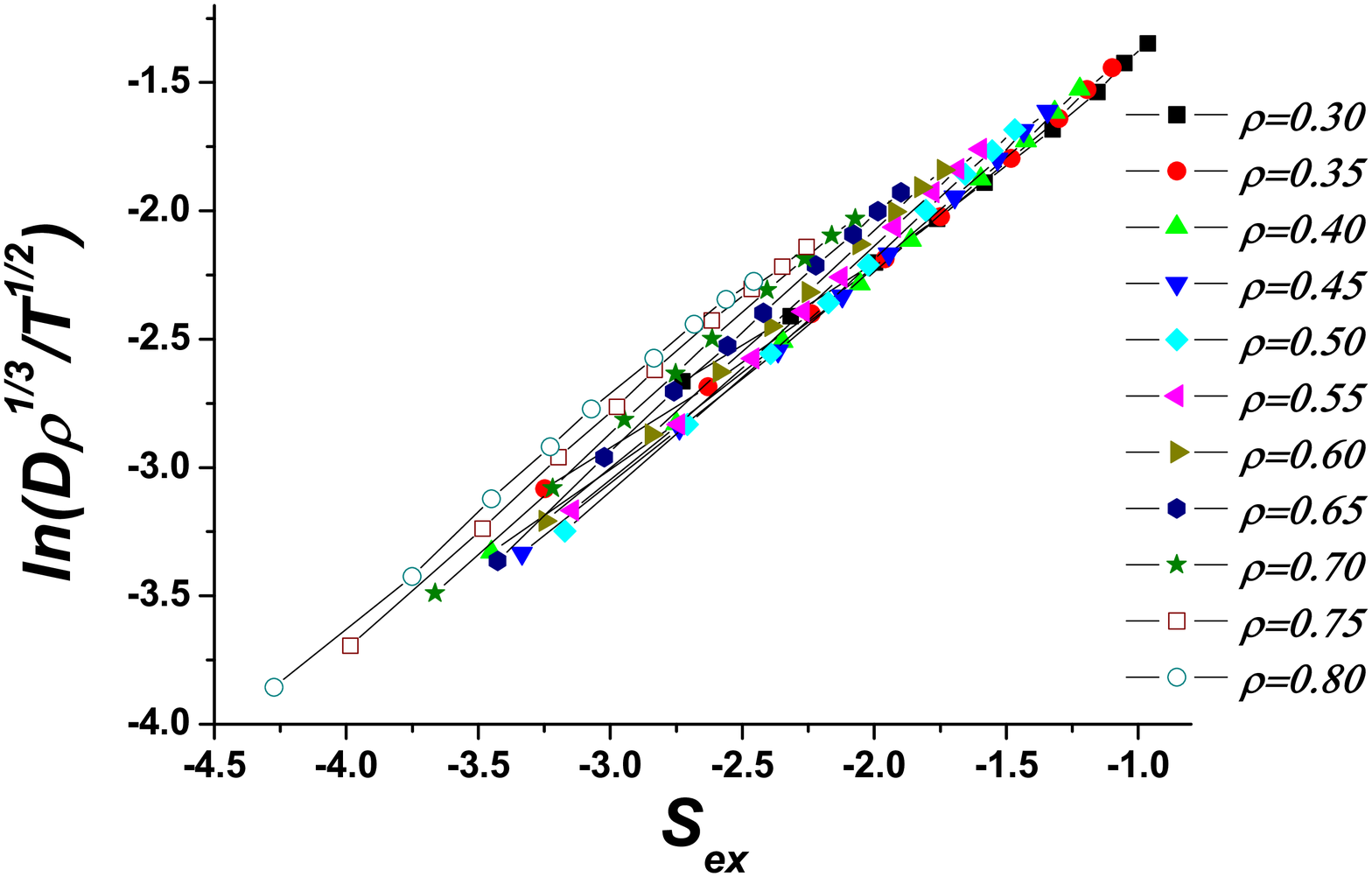}%

\includegraphics[width=6cm, height=6cm]{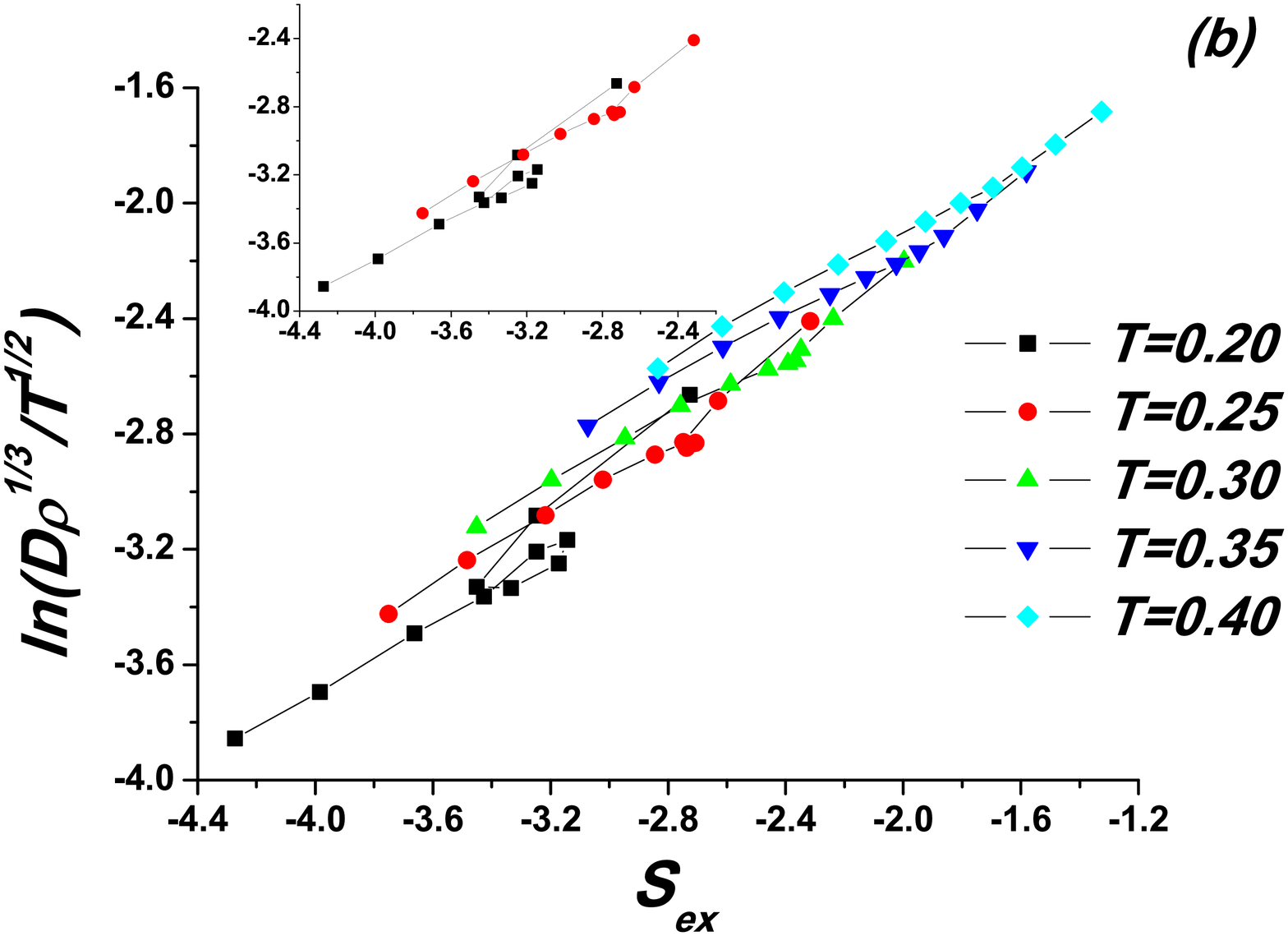}%
\caption{\label{fig:fig6} (Color online). (a) Reduced diffusion
logarithm for RSS along a set of isochors. (b) The logarithm of
the reduced diffusion coefficient as a function of $S_{ex}$ along
a set of isotherms for RSS. }
\end{figure}

In our previous publication \cite{weros} we showed that for this
system the Rosenfeld scaling relation along isotherms is not
applicable: the curves demonstrate the self crossing loops (see
Fig.~\ref{fig:fig6} (b)). Fig.~\ref{fig:fig6} (a) represents the
logarithm of the reduced diffusion coefficient along a set of
isochors for the RSS. As it can be seen from this figure, the
dependence of $D^*$ on $S_{ex}$ is linear. However, the slope of
the line shows an isochor dependence. Fig.~\ref{fig:fig6} (a)
shows that the slope remains approximately constant for low
densities ($\rho=0.3-0.55$) while on increasing the density the
slope also increases. From Figs.~\ref{fig:fig6} ((a) and (b)) one
can conclude that the Rosenfeld scaling relation hold along the
isochores, but breaks down along the isotherms.

\begin{figure}
\includegraphics[width=5cm, height=4cm]{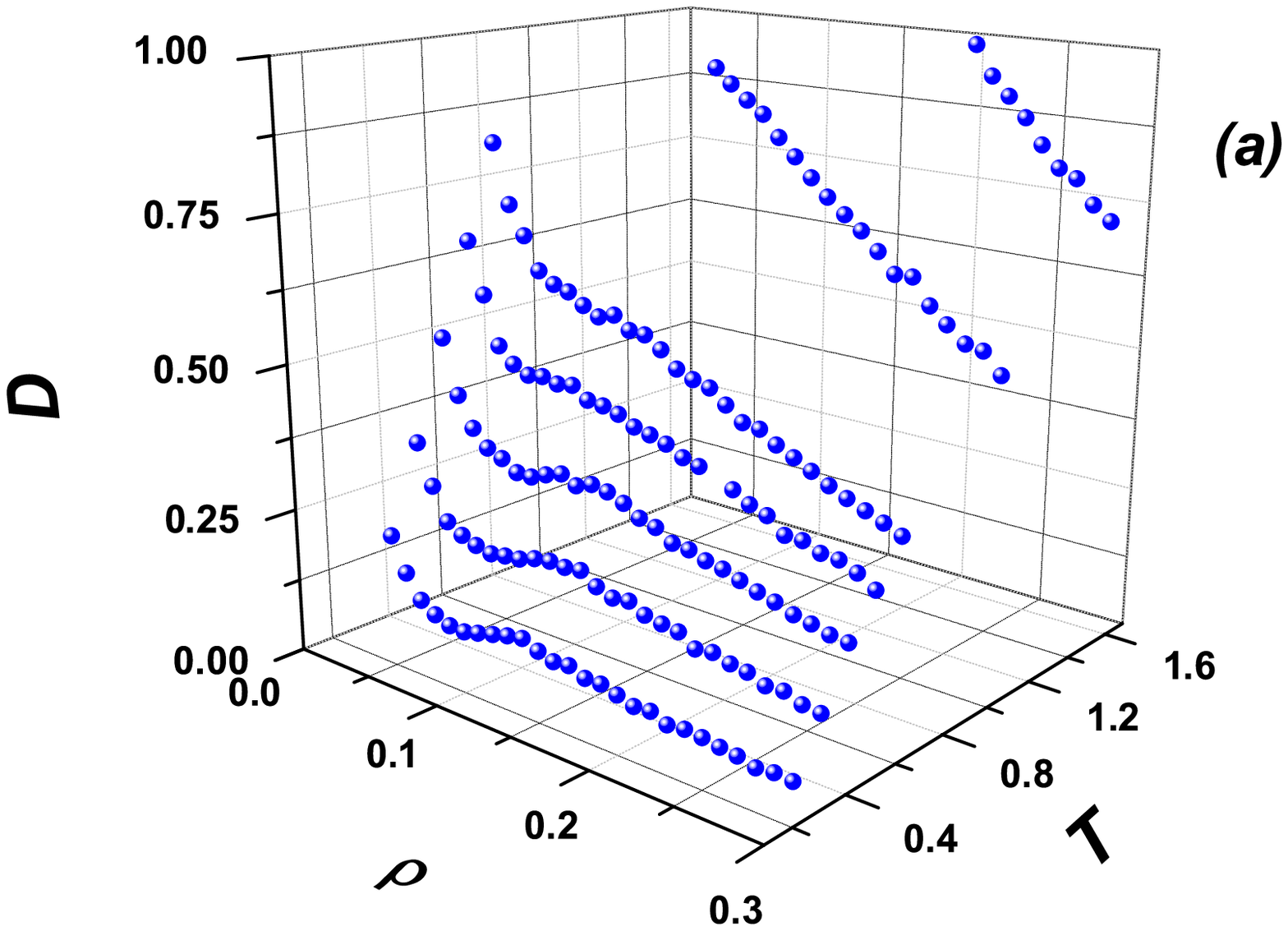}%
\bigskip
\bigskip

\includegraphics[width=5cm, height=4cm]{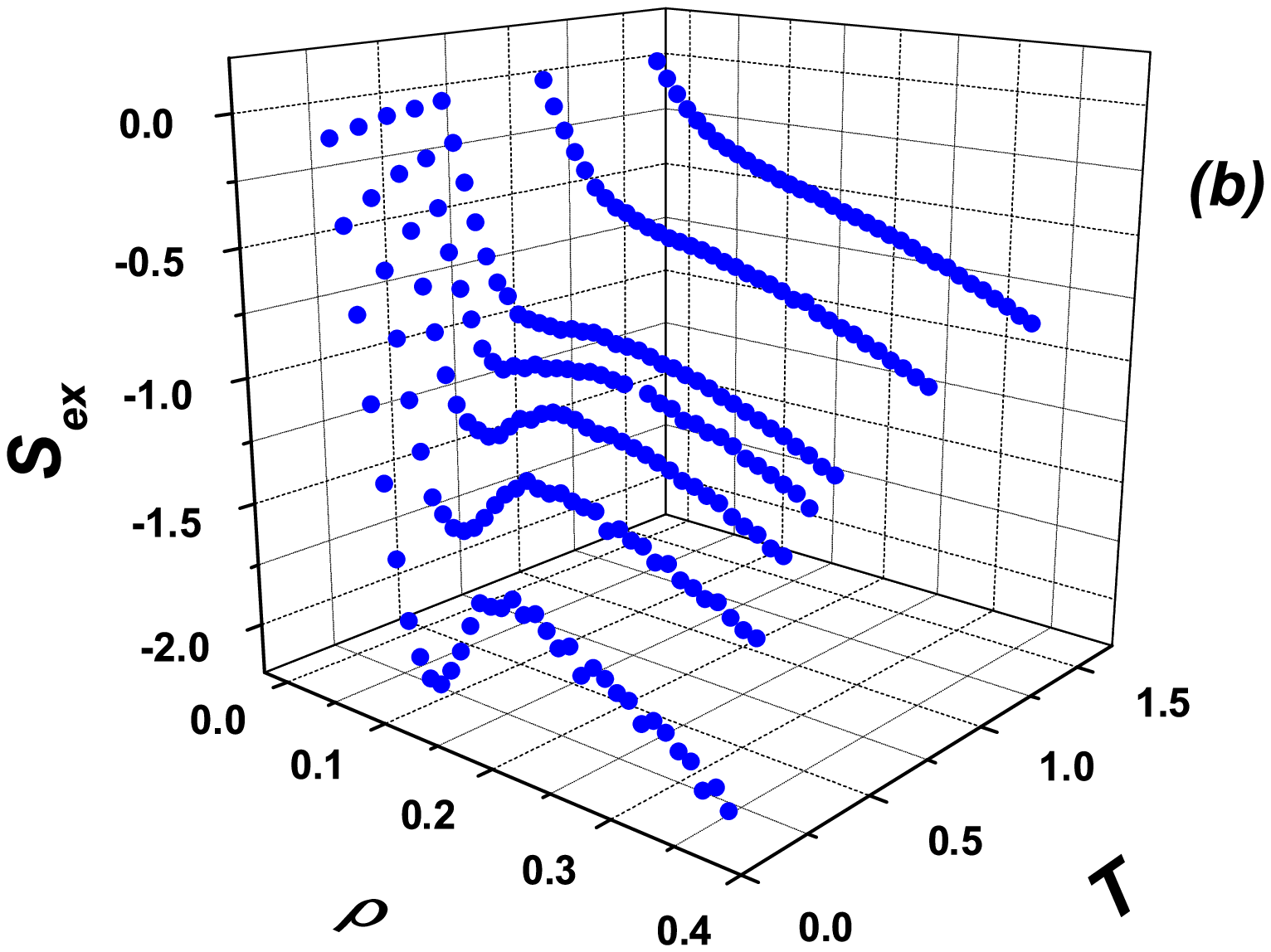}%
\bigskip

\caption{\label{fig:fig10} (Color online). (a) The diffusion
coefficient of the LJG system as a function of $\rho$ and $T$. (b)
The excess entropy of the LJG system as a function of $\rho$ and
$T$. }
\end{figure}

\begin{figure}
\includegraphics[width=6cm, height=6cm]{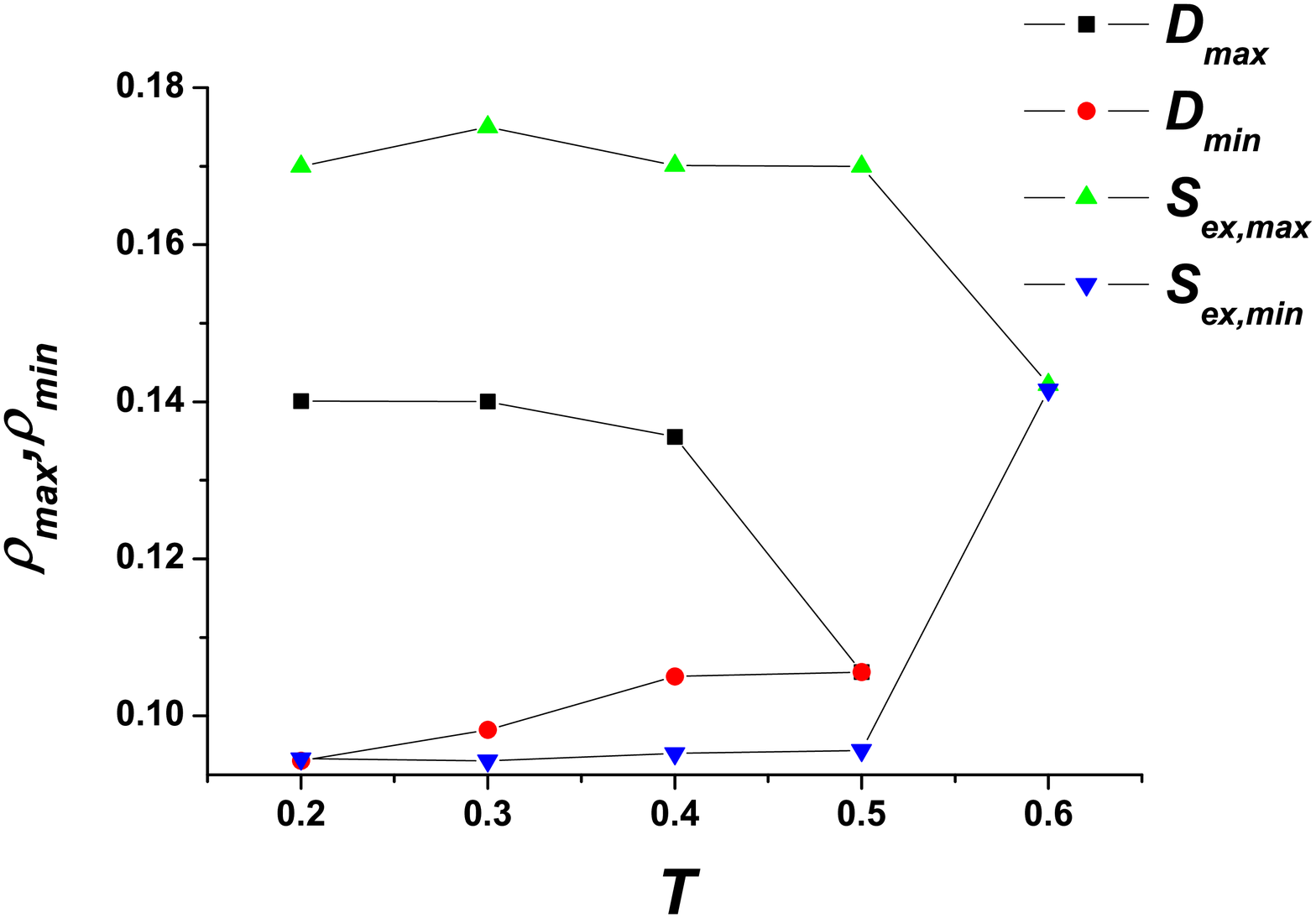}%

\caption{\label{fig:fig2}  The densities of maximum and minimum of
the diffusivity and the excess entropy for the several isotherms.}
\end{figure}

In order to show that the dependence of the water-like anomalies
and the Rosenfeld scaling relations on the trajectory in the
($\rho-T$) plane is a general property of the core-softened
systems we study another core softened potential introduced above
- the LJG potential. Fig.~\ref{fig:fig10}(a) presents the
diffusion coefficient of the LJG system as a function of $\rho$
and $T$. As in the case of RSS, one can see that at low
temperatures there is the diffusion anomaly.
Fig.~\ref{fig:fig10}(b) demonstrates the excess entropy as a
function of $\rho$ and $T$. From Figs.~\ref{fig:fig10}((a) and
(b)) one can see that there is a diffusion anomaly along the
isotherms, but there is no anomaly along the isochores (compare
Figs.~\ref{fig:fig1} ((a) and (b)). From these figures it follows
that the diffusion coefficient and the excess entropy along the
isotherms have similar qualitative behavior. However, the location
of the extremum points of diffusivity and excess entropy along the
isotherms is different (Fig.~\ref{fig:fig2}). It means that there
are some regions where one function increases while another one
decreases and vice versa. Clearly, this kind of behavior can not
be consistent with the Rosenfeld scaling formula. From this one
can conclude that the Rosenfeld scaling relation is not applicable
to the diffusivity along an isotherm for the LJG model. It should
be noted that the behavior depicted here is consistent with
simulation results for other water-like fluids (see, for example
Ref. \onlinecite{errington1}) therefore one can expect that the
violation of the Rosenfeld scaling relation along the isotherms is
the general feature for the systems with the water-like anomalies.

\begin{figure}
\includegraphics[width=6cm, height=6cm]{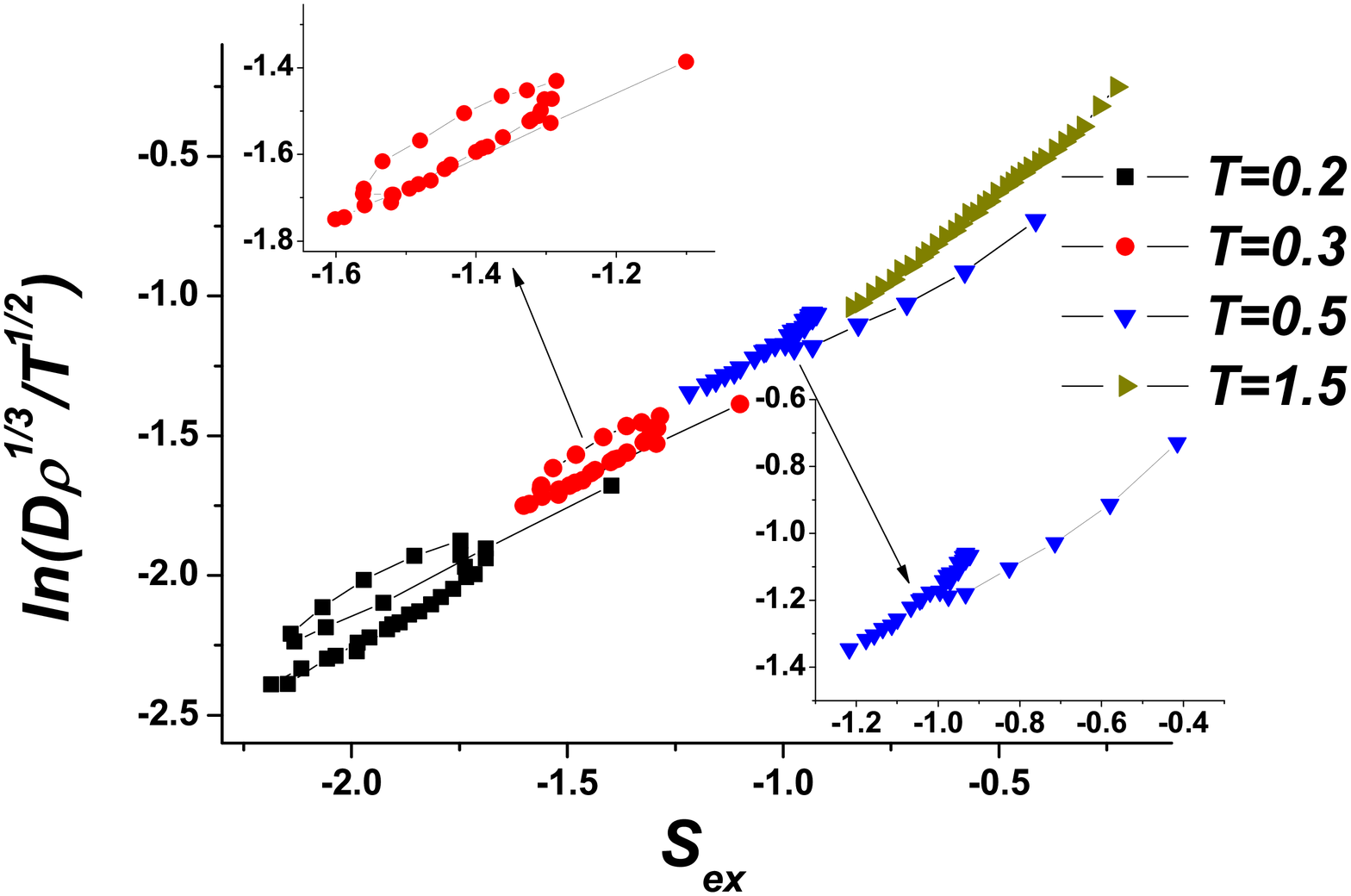}%

\includegraphics[width=6cm, height=6cm]{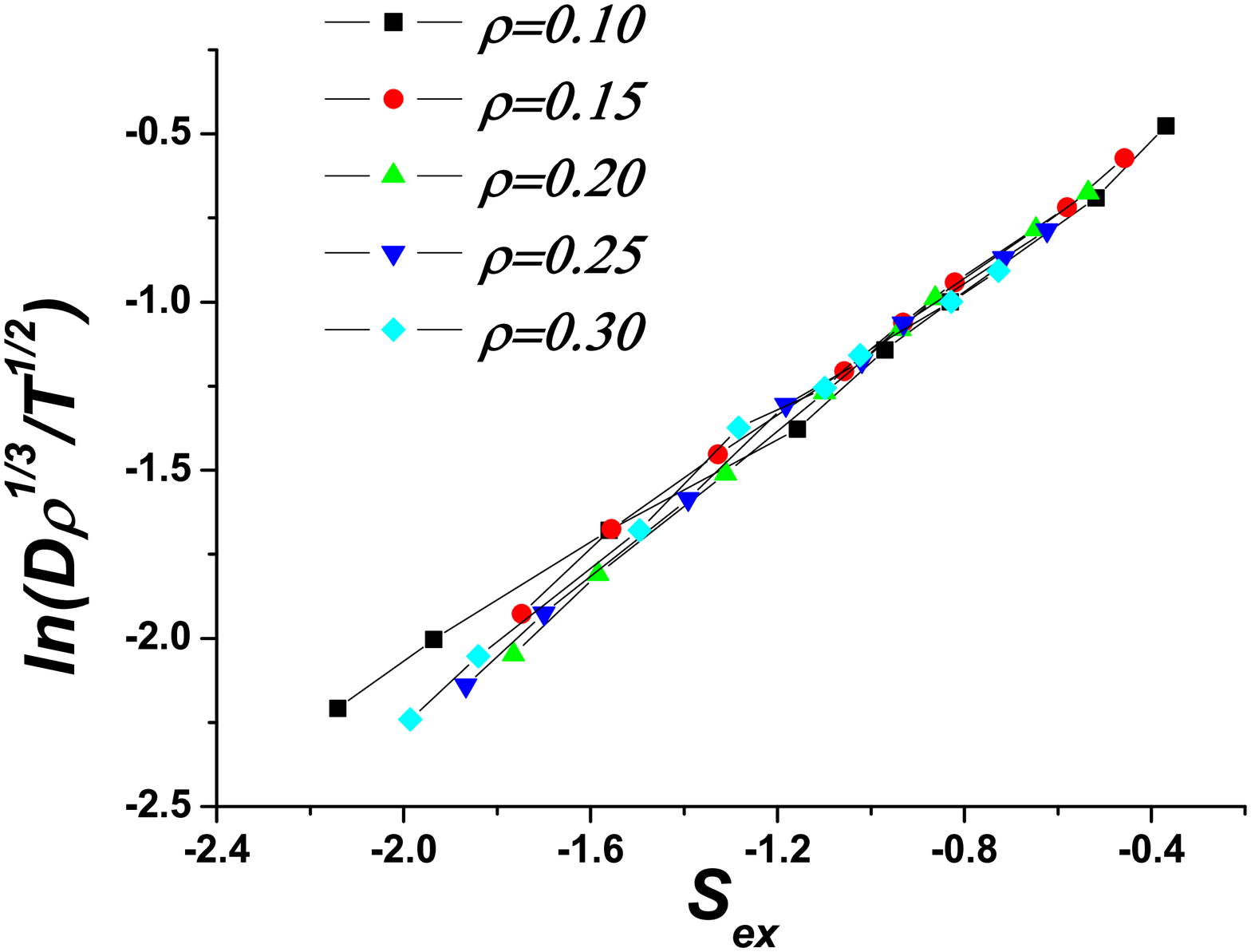}%

\caption{\label{fig:fig3} (Color online). (a) The logarithm of the
reduced diffusion coefficient for the LJG model along the several
isotherms. The insets correspond to the temperatures $T=0.3$
(upper panel) and $T=0.5$ (lower panel) in the larger scale. (b)
The logarithm of the reduced diffusion coefficient for the LJG
model along the several isochors.}
\end{figure}

Fig.~\ref{fig:fig3} (a) presents the logarithm of reduced
diffusion coefficient versus excess entropy along a set of
isotherms. One can see that at low temperatures the curves
demonstrate the selfcrossing loops like the ones observed for the
RSS model (see Ref. \cite{weros} anf Fig.~\ref{fig:fig6} (b)).
These loops become less pronounced with increasing the temperature
and at $T=0.5$ the self crossing disappears.

It can be seen from Figs.~\ref{fig:fig10}((a) and (b)), that both
the diffusivity and the excess entropy along isochors are
monotonous. It allows to expect that, as in the case of RSS, the
exponential relation between the diffusion coefficient and the
excess entropy holds along isochors. Fig.~\ref{fig:fig3} (b) shows
the $\ln(D^*)$ versus $S_{ex}$ along a set of isochors. From this
figure one can see that all curves with good accuracy correspond
to the Rosenfeld scaling relation. This result is in agreement
with Refs. \cite{errington2,indiabarb}. However,
Fig.~\ref{fig:fig3} (a) contradicts to the Refs.
\cite{errington2,indiabarb}. For example, in Fig. 3 of Ref.
\cite{indiabarb} the authors present the anomalous behavior of the
diffusion coefficient along the isotherms, but show the Rosenfeld
relation along the isochores and make the wrong conclusion that
the Rosenfeld-type excess entropy scaling is valid for the LJG
system.

In conclusion, in the present article we carry out a molecular
dynamics study of the two core-softened systems (RSS and LJG) and
show that the existence of the water-like anomalies in kinetic
coefficients in these systems depends on the trajectory in
$\rho-T$ plane along which the kinetic coefficients are
calculated. In particular, it is shown that the diffusion anomaly
does exist along the isotherms, but disappears along the isochores
and isobars. We analyze the applicability of the Rosenfeld entropy
scaling relations to these systems in the regions with the
water-like anomalies. It is shown that the validity of the of
Rosenfeld scaling relation for the diffusion coefficient also
depends on the trajectory in the $\rho-T$ plane along which the
kinetic coefficients and the excess entropy are calculated. In
particular, it is valid along isochors, but it breaks down along
isotherms.

\begin{acknowledgments}
We thank V. V. Brazhkin and Daan Frenkel for stimulating
discussions. Y.F. also thanks the Russian Scientific Center
Kurchatov Institute for computational facilities. The work was
supported in part by the Russian Foundation for Basic Research
(Grants No 08-02-00781 and No 10-02-00700) and Russian Federal
Program 02.740.11.5160.
\end{acknowledgments}


\end{document}